\documentclass[%
 aip,
 amsmath,amssymb,
 reprint,%
]{revtex4-1}

\usepackage{graphicx}
\usepackage{subfigure}
\usepackage{dcolumn}
\usepackage{bm}

\usepackage[utf8]{inputenc}
\usepackage[T1]{fontenc}
\usepackage{mathptmx}
\usepackage{amssymb}
\usepackage{epstopdf}
\usepackage[none]{hyphenat}

\usepackage[breaklinks=true,
colorlinks=true,
linkcolor=blue,
citecolor=blue,
urlcolor=blue]{hyperref}

\begin{document}

 \title{Multi-dimensional Vlasov simulations on trapping-induced sidebands of Langmuir waves}

  \author{Y. Chen}
\affiliation{Key Laboratory for Micro-/Nano-Optoelectronic Devices of Ministry of Education, School of Physics and Electronics, Hunan University, Changsha, 410082, China}
\affiliation{Institute of Applied Physics and Computational Mathematics, Beijing, 100094, China}

\author{C. Y. Zheng}
\affiliation{Institute of Applied Physics and Computational Mathematics, Beijing, 100094, China}
\affiliation{\mbox{HEDPS, Center for Applied Physics and Technology, Peking University, Beijing 100871, China}}
\affiliation{\mbox{Collaborative Innovation Center of IFSA (CICIFSA), Shanghai Jiao Tong University, Shanghai 200240, China}}

\author{Z. J. Liu}
\affiliation{Institute of Applied Physics and Computational Mathematics, Beijing, 100094, China}
\affiliation{\mbox{HEDPS, Center for Applied Physics and Technology, Peking University, Beijing 100871, China}}

\author{L. H. Cao}
\affiliation{Institute of Applied Physics and Computational Mathematics, Beijing, 100094, China}
\affiliation{\mbox{HEDPS, Center for Applied Physics and Technology, Peking University, Beijing 100871, China}}
\affiliation{\mbox{Collaborative Innovation Center of IFSA (CICIFSA), Shanghai Jiao Tong University, Shanghai 200240, China}}

\author{C. Z. Xiao } 
  \email{xiaocz@hnu.edu.cn}
  \affiliation{Key Laboratory for Micro-/Nano-Optoelectronic Devices of Ministry of Education, School of Physics and Electronics, Hunan University, Changsha, 410082, China}
  \affiliation{\mbox{Collaborative Innovation Center of IFSA (CICIFSA), Shanghai Jiao Tong University, Shanghai 200240, China}}
\date{\today}

  \begin{abstract}

   Temporal evolution of Langmuir waves is presented with two-dimensional electrostatic Vlasov simulations. In a muti-wavelength system, trapped electrons can generate sidebands including longitudinal, transverse and oblique sidebands. We demonstrated that oblique sidebands are important decay channels of Langmuir waves, and the growth rate of oblique sideband is smaller than the longitudinal sideband but higher than the transverse sideband. When the amplitudes of sidebands are comparable with that of Langmuir wave, vortex merging occurs following the broadening of longitudinal and transverse wavenumbers, and finally the system is developed into a turbulent state. In addition, the growth of sidebands can be depicted by the nonlinear Schr\"{o}dinger model (Dewar-Rose-Yin (DRY) model) with non-Maxwellian Landau dampings. It shows the significance of particle-trapping induced nonlinear frequency shift in the evolution and qualitative agreement with Vlasov simulations.
  \end{abstract}


  \maketitle

\section{Introduction}\label{introduction}

In common electron-ion plasmas, some collective modes of interaction or eigenmodes can be found. And these waves are related to the charge density fluctuations. A low frequency one called ion acoustic wave corresponds to the characteristic frequency determined by ions and a high frequency one called Langmuir wave\cite{kruer,Nicholson} corresponds to the characteristic frequency determined by electrons. In inertial confinement fusion (ICF),\cite{ICF1,I_D,Tik} an incident pump laser can easily decay to a reflected electromagnetic wave and a Langmuir wave, which is the so-called stimulated Raman scattering (SRS).\cite{chen,chen1,Brunner1,xiao1,xiao2,xiao3,xiao4,xiao5} Because this instability consumes pump energy and produces hot electrons, which are detrimental to fusion, it is crucial to find the saturation mechanism of SRS.\cite{Estabrook,Vu,yin,Lin}

At $1960s$, Kruer $et$ $al.$ revealed a trapped-particle instability that Langmuir wave can decay to sidebands,\cite{KDS} and it can serve as a saturation mechanism of SRS.\cite{Brunner1}  The so-called Kruer, Dawson, and Sudan (KDS) model achieves great success in explaining the longitudinal sidebands in one dimension.\cite{Brunner2,Krasovsky} Recently, Friou et al. studied sidebands of Langmuir waves as a saturation of SRS by using Vlasov and particle in cell (PIC) code in one dimensional geometry,\cite{Friou} the results from different codes agree with each other and they found that the amplitude of growth rates of longitudinal sidebands have scaling law $\gamma \sim g_{1}\phi^{g_{2}}$, where $g_{2} \sim 0.6 - 0.9$ for different wavenumber of Langmuir waves. Also, another theoretical model given by Dewar, Kruer and Manheimer is used to study the sideband instability.\cite{Dewar} This model is based on the nonlinear Schr\"{o}dinger equation and shows that the sidebands are originated from the nonlinear frequency shift by trapped electrons.\cite{Berger1} Dewar's model was also be extended to two dimensions (so called Dewar-Rose-Yin model, $i.e.$ DRY model) to analyze the filamentation of Langmuir wave.\cite{rose1,rose2,winjum}

Recently, the transverse decay channels of plasma waves get many attentions.\cite{Berger2,Denis1,Denis2,Chapman}  Berger $et$ $al.$  used both generalization of KDS model (GKDS) and DRY model to study the transverse sidebands of Langmuir wave.\cite{Berger2} The growth rates of transverse sidebands of Langmuir wave are obtained from these two models and reasonably agreed with their single-wavelength Vlasov simulations. They found that transverse sidebands can lead to the filamentation of Langmuir wave.  Also, D. A. Silantyev $et$ $al.$ studied the transverse sideband of Langmuir waves.\cite{Denis1,Denis2} They found the scaling of the maximum growth rate is $\gamma_{max}\propto\sqrt{\phi}$ by using Vlasov-Possion simulations, where $\phi$ is the electrostatic potential of Langmuir waves.   Similarly, Chapman $et$ $al.$ observed that the ion acoustic waves (IAW) have the longitudinal decay channel, two-ion decay (TID), and the transverse decay channel, off-axis instability (OAI). Although they failed to explain OAI by existing theories, they conjectured the OAI is an ion-driven trapped particle instability.\cite{Chapman} We should note that the nature of TID instability is different from sideband instability.  These  works indicate that trapped particles can generate both longitudinal and transverse instabilities. Inspired by the transverse modulation of plasma waves, we believe that sidebands of Langmuir wave is a muti-dimensional instability, and the oblique sidebands whose wavevector has an oblique angle to the wavevector of Langmuir wave should be considered. In Berger's Vlasov simulations, they only considered one longitudinal wavelength of Langmuir wave, thus they cannot observe the longitudinal and oblique sidebands.\cite{Berger2} Now, we fill this gap by using multi-wavelength Vlasov simulations and DRY model to study the sidebands of Langmuir waves in multi-dimensions .

In this paper, first, we use two-dimensional Vlasov-Poisson simulations to observe the nonlinear evolution of sidebands of Langmuir waves, including longitudinal sidebands, oblique sidebands and transverse bands. We should note that the oblique sidebands are just the trapped particle instability in a new direction. We find that oblique sidebands are important decay modes of Langmuir waves, which is similar to longitudinal sidebands and transverse sidebands. When the amplitudes of sidebands are comparable with that of Langmuir wave, sidebands saturate through a violent process featured as vortex merging in the phase space. This phenomenon is firstly discovered  by Brunner $et$ $at$.\cite{Brunner2} Recently, Yang $et$ $al.$ also studied vortex merging by Vlasov simulations.\cite{yang} The $k_{x}$, $k_{y}$  component of modes then broaden rapidly. Finally, the system is developed into a turbulent state lasting from $t\omega_{pe}=880$ to the end of simulations. Second, we extent the DRY model to two dimension and obtain the growth rates of sidebands. The Landau damping of Langmuir waves in a non-Maxwellian distribution is considered.  In simulations, we obtain the scaling law for longitudinal sidebands is $\gamma \sim \phi^{0.75}$, and that for oblique sideband with $k_{y} = 0.03125$ is $\gamma \sim \phi^{0.85}$. The $g_{2}$ for longitudinal sidebands in our paper agree with the results in Friou's work,\cite{Friou} $g_{2} \sim 0.6-0.9$.  In our theoretical results, the scaling law for longitudinal sidebands is $\gamma \sim \phi^{0.5}$, and that for oblique sideband with $k_{y} = 0.03$ is $\gamma \sim \phi^{0.66}$. Our theoretical model can  describe the longitudinal sidebands, oblique sidebands, and transverse sidebands. The growth rates of sidebands qualitatively agree with our Vlasov simulation results.

This paper is structured in the following ways. Firstly, in Sec.~\ref{Simulation model}, we use Vlasov simulations to study the nonlinear evolution of the sidebands of Langmuir waves in two dimensions. Secondly,  DRY model are performed to study the growth rate of sidebands in Sec.~\ref{theory model}. At last, the conclusion and discussion are shown in  Sec.~\ref{conclusion}.

\section{Two dimensional Vlasov simulations }\label{Simulation model}

\subsection{Excitation of Langmuir waves}
The evolution of Langmuir waves is governed by Vlasov-Poisson equations. In two-dimensional (2D) and multi-wavelength system, the Vlasov-Poisson equations are given by \begin{equation}
\begin{split}
&\frac{\partial f_{e}}{\partial t} + v_{x}\frac{\partial f_{e}}{\partial x}+v_{y}\frac{\partial f_{e}}{\partial y}-\frac{eE_{x}}{m_{e}}\frac{\partial f_{e}}{\partial v_{x}}-\frac{eE_{y}}{m_{e}}\frac{\partial f_{e}}{\partial v_{y}}=0,\\
&\frac{\partial^{2}\phi}{\partial x^{2}}+\frac{\partial^{2}\phi}{\partial y^{2}}=\rho\\
& E_{x} = -\frac{\partial \phi}{\partial x}+E_{ext},E_{y} = -\frac{\partial \phi}{\partial y},
\end{split}
\end{equation}
where $f_{e}$ is the distribution function, $E_{x}$ and $E_{y}$ are the electrostatic field at longitudinal and transverse direction, respectively, $\phi$ is electrostatic  potential, and $\rho$ is the charge density. $E_{ext}$ is an external electric field, longitudinally driving the Langmuir wave. Ion motion is not considered in our paper since we mainly focus on the Langmuir wave.

\begin{figure}[tbp]
    \begin{center}
      \includegraphics[width=0.5\textwidth,clip,angle=0]{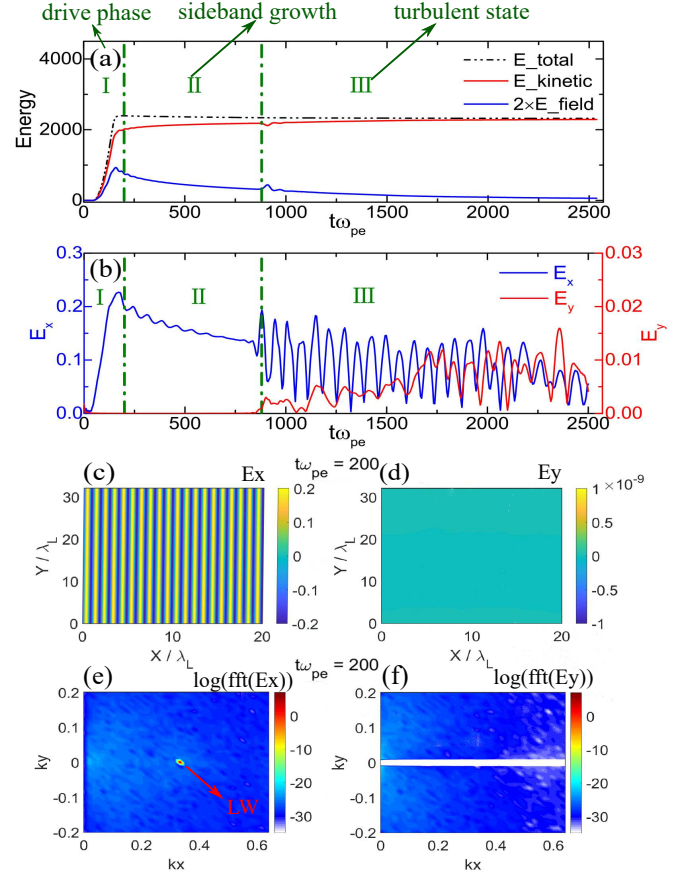}
      \caption{\label{energy}$k_{L}\lambda_{De} = 1/3$, the spatial space is $L_{x} = 20 \lambda_{L}$, $L_{y} = 32 \lambda_{L}$. (a) The temporal evolution of field energy $E_{field}$ (blue line), electron kinetic energy $E_{kinetic}$ (red line) and the total energy $E_{total} = E_{field}+E_{kinetic}$ (black dashed line). The field energy times $2$ for better view. (b) is the temporal evolution of envelope of $E_{x}$ and $E_{y}$ at the center of the simulation box. (c) and (d) are the $E_{x}$ and $E_{y}$ at $t = 200 \omega_{pe}^{-1}$.  (e) and (f) are the corresponding spectrum at $k_{x}$-$k_{y}$ space. }
    \end{center}
  \end{figure}

We have developed a 2D version of Vlasov-Poisson code, \emph{PLAW2d} (2D Plasma Wave simulation), which uses Van Leer scheme (VL3)\cite{VL,VL2} to solve the Vlasov equation and fast Fourier transform (FFT) to solve the Poisson equation \cite{yang,yangs,feng}. The spatial domain occupies $X\in[0,L_{x}]$ and $Y\in[0,L_{y}]$, and the boundary conditions are periodic at each side. The  lengths of simulation box are $L_{x} = 20 \lambda_{L}$ and $L_{y} = 32 \lambda_{L}$, where $\lambda_{L} = 2\pi/k_{L}$ is the wavelength of Langmuir wave and $k_{L} = 1/3 \lambda_{De}^{-1}$ is the corresponding wavenumber. We choose $k_{L} = 1/3 \lambda_{De}^{-1}$ as the wavenumber in simulation because it corresponds to the kinetic regime of SRS in ICF. Some nonlinear effects by particle trapping will occur in kinetic regime, such as trapped particle instability (TPI), inflation of SRS because of the decreasing of Landau damping of Langmuir wave, and spatially auto-resonant Stimulated Raman Scattering due to the nonlinear frequency shift in inhomogeneous plasma. The physical spaces are discretized by $\Delta x = 1.0471 \lambda_{De}$ and $\Delta y = 6.032 \lambda_{De}$, where $\lambda_{De}$ is the electron Debye length. The velocity domain is $v_{x,y}\in [-12v_{te},12v_{te}]$ and they are discretized by $\Delta v_{x} = \Delta v_{y} = 0.1v_{te}$, where $v_{te} = \sqrt{T_{e}/m_{e}}$ is the thermal velocity of electrons, $T_{e}$ and $m_{e}$ are the temperature and mass of electrons, respectively. Therefore, the four dimensional meshgrids are  $360\times100\times241\times241$. Electrons in plasma are initialized by 2D Maxwellian distribution function, $f_{e0}(v_{x},v_{y}) = \frac{1}{2\pi v_{te}} exp(-\frac{v_{x}^{2}+v_{y}^{2}}{2v_{te}^{2}})$. The total simulation time is $t_{total} = 2500 \omega_{pe}^{-1}$, and the time step is $dt = 0.1 \omega_{pe}^{-1}$, where $\omega_{pe} = \sqrt{4\pi n_{e}e^{2}/m_{e}}$ is the plasma frequency and $n_{e}$ is the electron density of plasma. We should notice that in the remaining paper the units are normalized to electron unit, such as, $v_{te} =1$, $\omega_{pe} = 1$ and $\lambda_{De} = 1$, etc.

  At the drive phase shown in Fig.~\ref{energy}(a) , $t = 0\sim200 \omega_{pe}^{-1}$,  an external  electrostatic field
  \begin{equation}
  E_{ext} = E_{ext}^{max}\left[1+(\frac{t-\tau}{\triangle\tau})^{n}\right]^{-1}sin(k_{L}x-\omega_{L}t),
 \end{equation}
 is used to induce the Langmuir waves, where $E_{ext}^{max}= 0.01 m_{e}\omega_{pe}v_{te}/e$ is the amplitude of pump drive, $\tau = 100\omega_{pe}^{-1}$, $\triangle\tau = 50\omega_{pe}^{-1}$, $n = 10$ and $\omega_{L} = \sqrt{\omega_{pe}^{2}+3k_{L}^{2}v_{te}^{2}}$ is the Bohm-Gross dispersion relation of Langmuir wave and the phase velocity is $v_{\phi}=\omega_{L}/k_{L}$. In stage \uppercase\expandafter{\romannumeral1} shown in Fig.~\ref{energy}(a), the field energy and electron kinetic energy increase exponentially because the Langmuir waves are resonant excited.  Fig.~\ref{energy}(b) is the time history of $E_{x}$ and $E_{y}$ envelopes at the center of the simulation box, and it shows the amplitude of $E_{x}$ increases to $0.2$ at $t = 200 \omega_{pe}^{-1}$. Fig.~\ref{energy}(c) to (f) are $E_{x}$ and $E_{y}$ in physical space and the their spectra in $k_x-k_y$ space at $t = 200 \omega_{pe}^{-1}$, respectively. In Fig.~\ref{energy}(c), there occupies the longitudinal Langmuir wave with $20$ wavelengths, but no fields are observed in the transverse direction as shown in Fig.~\ref{energy}(d). Correspondingly, the Langmuir wave (LW) mode are clearly excited at $k_{x} = 1/3, k_{y} = 0$ in Fig.~\ref{energy}(e), and no mode are shown in the transverse directions in Fig.~\ref{energy}(f).

\subsection{Growth of sidebands}
\begin{figure}[tbp]
    \begin{center}
      \includegraphics[width=0.5\textwidth,clip,angle=0]{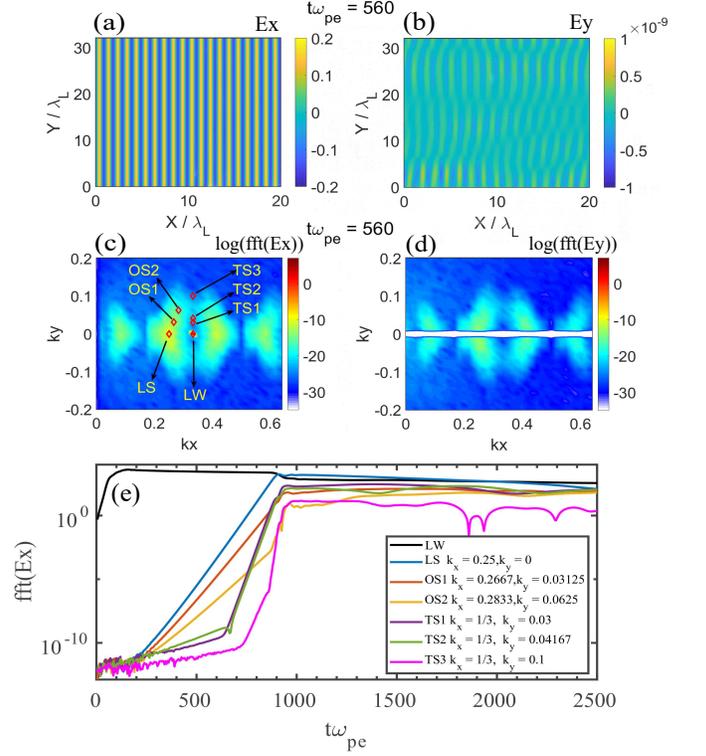}
      \caption{\label{sideband}(a) and (b) are the $E_{x}$ and $E_{y}$ at $t = 560 \omega_{pe}^{-1}$.  (c) and (d) are the corresponding spectrum at $k_{x}$-$k_{y}$ space. In (c), the red diamonds stand for the selected sidebands of Langmuir waves. (e) The evolution of Langmuir wave (black line), the growth of longitudinal sideband(blue line), oblique sidebands (red and orange lines) and transverse sidebands(violet, green and manganese purple lines).}
    \end{center}
  \end{figure}

  At $t = 200 \omega_{pe}^{-1}$, the external field is turned off, leaving the Langmuir waves propagating freely. The excited wave is, however, unstable to many instabilities, such as sideband instability, modulational instability, and filamentation instability etc. The trapped-particle-induced sidebands of Langmuir wave are the earliest-developing ones. In Fig.~\ref{energy}(a), the stage \uppercase\expandafter{\romannumeral2} ($t = 200\sim880 \omega_{pe}^{-1}$) is defined as the growth of sidebands where three kinds of sidebands are observed. The first one is the longitudinal sideband, $i.e.$ $k_{x} \neq k_{L}, k_{y} = 0$. This kind of sideband is well-studied in one-dimensional systems.\cite{Brunner1,KDS,Brunner2,Dewar,yang} The second one is the transverse sideband defined by $k_{x} = k_{L}, k_{y} \neq 0$ and discussed by Berger et al. \cite{Berger2} The last kind is called the oblique sideband, $i.e.$ $k_{x} \neq k_{L}$ and $k_{y} \neq 0$, which exists in multi-wavelength and multi-dimensional systems and has never been discovered before.

  \begin{figure}
    \begin{minipage}[t]{0.5\linewidth}
    \centering
    \includegraphics[width=2.07in]{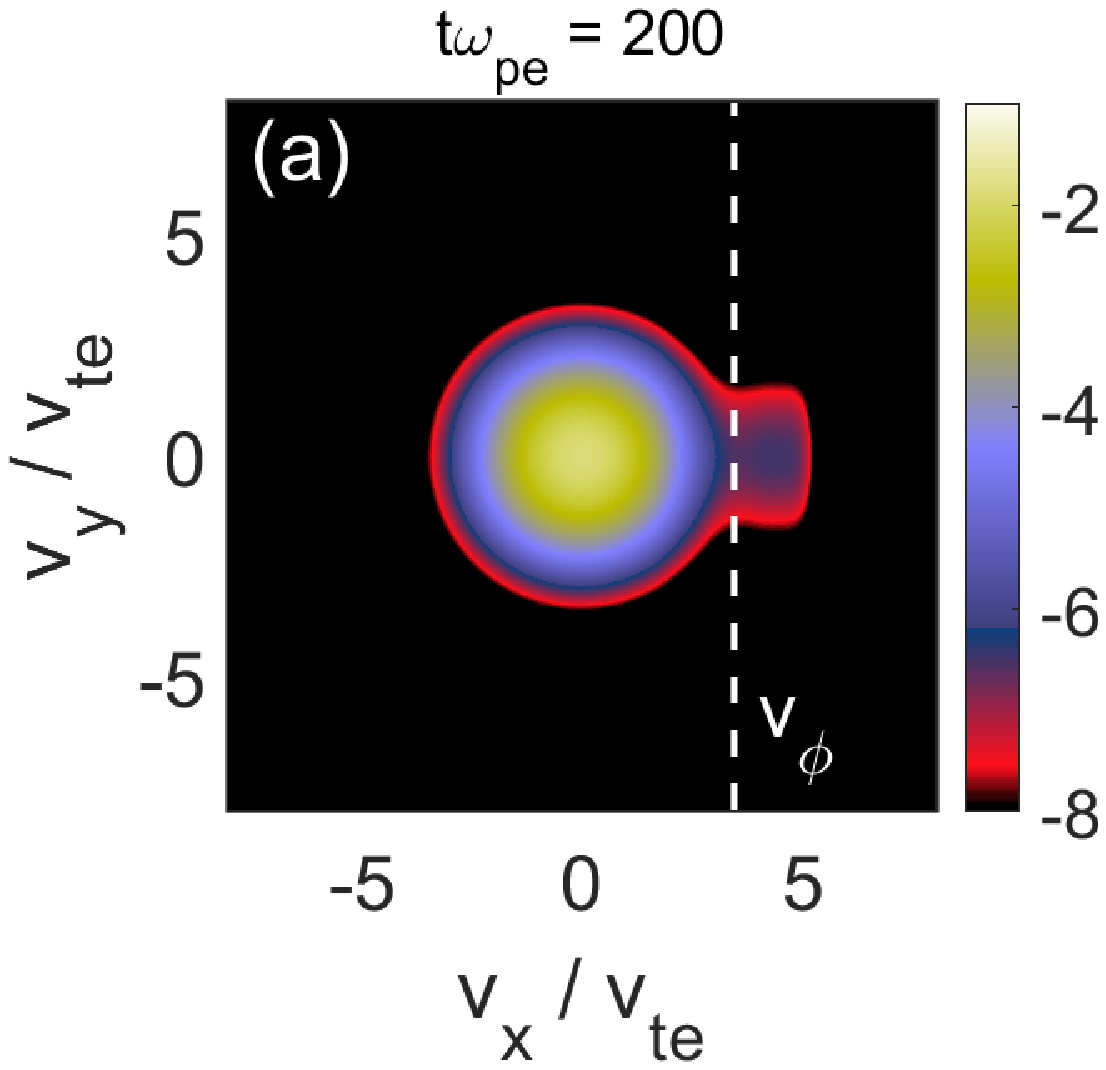}
    \label{fig:side:a}
    \end{minipage}%
    \begin{minipage}[t]{0.5\linewidth}
    \centering
    \includegraphics[width=2.07in]{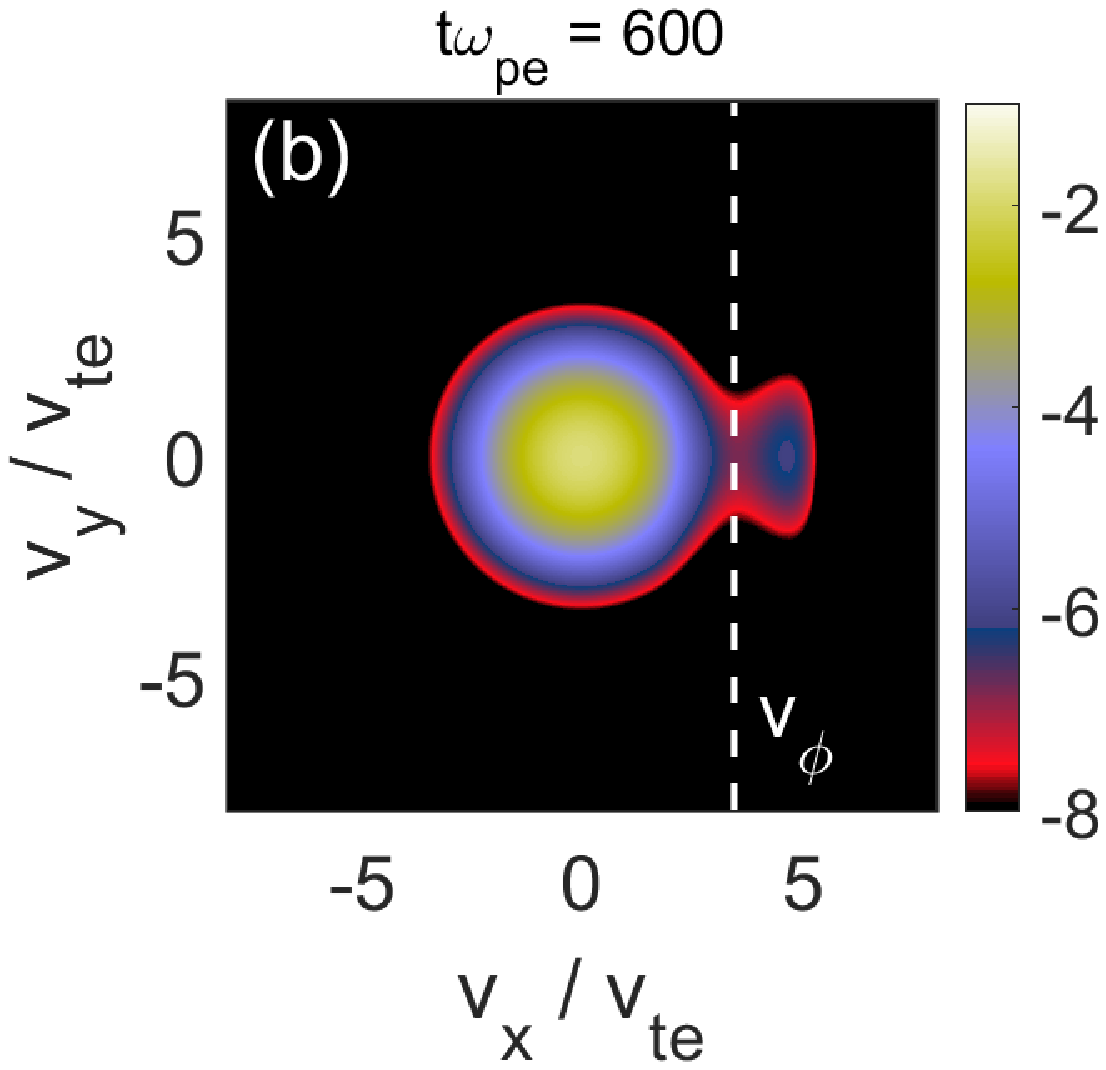}
    \label{fig:side:b}
    \end{minipage}
    \caption{\label{vx_vy}(a) Distribution function of electrons in $v_{x}$-$v_{y}$ space at $t\omega_{pe} = 200$. (b) Distribution function of electrons in $v_{x}$-$v_{y}$ space at $t\omega_{pe} = 600$.}
\end{figure}

  Fig.~\ref{sideband}(a) and (b) are $E_{x}$ and $E_{y}$ in the physical space at $t = 560 \omega_{pe}^{-1}$. In addition to the nearly unchanged longitudinal fields, modes with small amplitudes are observed in Fig.~\ref{sideband}(b), which imply sidebands with $k_{y} \neq 0$ are excited. Fig.~\ref{sideband}(c) and (d) are the corresponding spectra of sidebands in $k_{x}$-$k_{y}$ space where longitudinal sidebands, transverse sidebands and oblique sidebands emerge clearly. In Fig.~\ref{sideband} (c), we pick some modes for further analyses: LW represents the Langmuir wave at $k_{x} = 1/3, k_{y} = 0$; LS stands for the longitudinal sidebands with the maximum growth rate at $k_{x} = 0.25, k_{y} = 0$; OS1 and OS2 are two chosen oblique sidebands for $k_{x} = 0.2667, k_{y} = 0.03125$ and $k_{x} = 0.2833, k_{y} = 0.0625$; TS1, TS2 and TS3 are three transverse sidebands at $k_{x} = k_{L}$ and $k_{y} = 0.03, 0.04167$ and $0.1$, respectively.

  \begin{figure}[tbp]
    \begin{center}
      \includegraphics[width=0.48\textwidth,clip,angle=0]{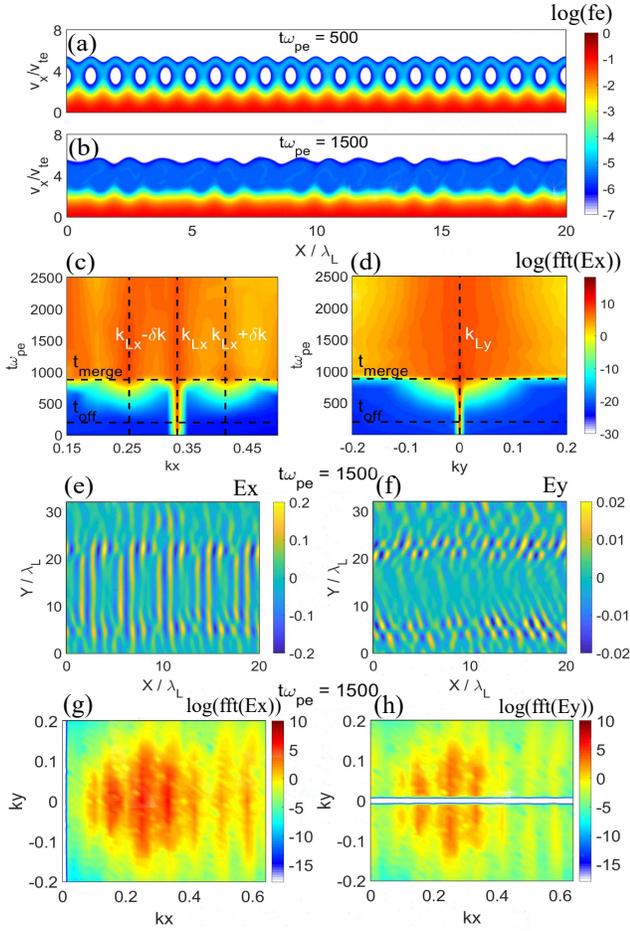}
      \caption{\label{vertex}(a) and (b) are the electron trapped structures at $t = 500 \omega_{pe}^{-1}$ and $t = 1500 \omega_{pe}^{-1}$. (c) and (d) are the temporal evolution of $k_{x}$ and $k_{y}$ spectrum. $t_{off} = 200 \omega_{pe}^{-1}$ is the time when the external driver is turned off, $t_{merge} = 880 \omega_{pe}^{-1}$ is when the vortex merging happens. (e) and (f) are the $E_{x}$ and $E_{y}$ at $t = 1500 \omega_{pe}^{-1}$.  (g) and (h) are the corresponding spectrum at $k_{x}$-$k_{y}$ space.}
    \end{center}
  \end{figure}

  The growth of these sidebands are shown in Fig.~\ref{sideband}(e). When resonant excited, the Langmuir wave (black line) gives its energy to other modes and particles. The fastest growing mode is the longitudinal sideband denoted by the blue line, and then it comes the oblique sidebands (the red line and the orange line), slowing down with the increase of $k_{y}$. The transverse sidebands are the slowest ones as indicated by the purple, green, and pink lines. As $k_y$ increases the growth rate of transverse sideband increases first and then decreases, showing the same features revealed by Berger et al. \cite{Berger2} The maximum growth rate of the transverse sideband is located at $k_{y}=0.04167$. However, this is valid only when $t = 200 \sim 600 \omega_{pe}^{-1}$, since a two-stage growth of transverse sidebands is observed. When $600\omega_{pe}^{-1}<t<1000\omega_{pe}^{-1}$, a rapid growth of the transverse sideband emerges, violating the linear growth pattern of sideband instabilities. Two-stage growth of sidebands can also be seen in earlier works.\cite{Brunner2,Berger2,Denis1} However, no excellent explanation of this phenomenon was provided by early works. Fig.~\ref{vx_vy} shows the phase space structures in $v_x$ and $v_y$ at two typical times $t = 200 \omega_{pe}^{-1}$ and $t = 600 \omega_{pe}^{-1}$. Based on the discovery of a recent work,\cite{yanxia} when the beam-like distribution function of electron was formed, there will be a beam plasma instability induced by the beam of electrons. The average beam velocity is equal to the phase velocity of Langmuir wave. This instability may lead to the two-stage growth of initially slower growing modes. Accordingly, the beam-like of distribution function of electron is also formed in Fig.~\ref{vx_vy}(b). So,we guess the two-stage growth of initially slower growing modes may be related to the beam-plasma instability.

  We summarize the results above that sidebands of the Langmuir wave can be excited by trapped electrons. In a two-dimensional and multi-wavelength system, there are sidebands in three kinds of directions, i.e. longitudinal sidebands, transverse sidebands and oblique sidebands. The maximum growth occurs with the longitudinal sideband, and the second maximum growth belongs to the oblique sideband, which is the trapping-induced sideband in a new direction. The slowest is transverse sideband.

  \subsection{Vortex merging and wave-particle-interaction-induced turbulence}

  At $t \approx 900 \omega_{pe}^{-1}$ in Fig.~\ref{energy}(a), there exists a violent energy exchange between kinetic and field energy, which implies strong wave-particle interactions. Also, in Fig.~\ref{energy}(b), $E_{x}$ and $E_{y}$ become unprecedentedly oscillative, revealing a whole new stage. This specific stage \uppercase\expandafter{\romannumeral3} is termed as the turbulent state. The term, turbulent state, means an utter destruction of the coherent phase-space structures into chaos, which is described in a recent one-dimensional study.\cite{Brunner2,yang}

 When different modes have similar amplitudes comparable to the fundamental Langmuir wave, they will interact with the BGK structure of Langmuir waves.\cite{Shoucr1,Shoucr2,Shoucr3} At $t \approx 880 \omega_{pe}^{-1}$, it hits the critical point and then vortex merging happens. In Fig.~\ref{vertex}(a), there are $20$ vortices in the phase space before vortex merging, but after that only $15$ vortices are left with irregular trajectories shown in Fig.~\ref{vertex}(b). The transition is demonstrated clearly in Fig.~\ref{vertex}(c) and (d), which plot the temporal evolutions of $E_{x}$ in $k_{x}$ and $k_{y}$ space, respectively. The black dashed lines, $k_{x} = 1/3$ in \ref{vertex}(c) and $k_{y}=0$ in \ref{vertex}(d)  represent the Langmuir wave. After turning off the driver, the longitudinal sidebands with the maximum growth rates occur along the black dashed lines ($k_{Lx} - \delta_{k}$ and $k_{Lx} + \delta_{k}$) in Fig.~\ref{vertex}(c). Oblique and transverse sidebands are subsidiary signals apart from the main Langmuir wave and longitudinal sidebands signals. An abrupt transition occurs at $t_{merge} = 880 \omega_{pe}^{-1}$, and within less than $50 \omega_{pe}^{-1}$ the spectrum explodes. After that transition, the broadening spectrum seems to be a long-term and stable structure, so we suppose the final state of the Langmuir wave is reached. Since the spectrum is much like the turbulence structure, we term the final state as the wave-particle-interaction-induced turbulent state.

  The turbulent states in physical space and spectral space are shown in Figs.~\ref{vertex}(e) to (h) at $t = 1500 \omega_{pe}^{-1}$. Filaments-like structures are easily observed in Fig.~\ref{vertex}(e) and (f). These structures are characterized by the stripes in the $k_x$-$k_y$ space resulting from nonlinear expansions of the sidebands along $k_y$.

\subsection{Amplitude dependence of sidebands}

\begin{figure}[tbp]
    \begin{center}
      \includegraphics[width=0.48\textwidth,clip,angle=0]{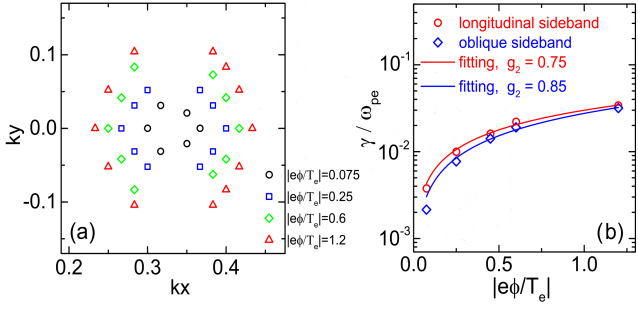}
      \caption{\label{gamma_s}Vlasov simulation results. (a) The positions of oblique sidebands for different amplitude of field at $k_{x}$-$k_{y}$ space. (b) The scaling law of growth rates of sidebands are obtained by using power law fitting $\gamma = g_{1}\phi^{g_{2}}$. We find that $g_{2} = 0.75$ for longitudinal sideband and $g_{2} = 0.85$ for oblique sideband with $k_{y} = 0.03125$.}
    \end{center}
  \end{figure}

By changing the starting amplitude of Langmuir wave potential, we find the nonlinear evolution experiencing such processes is rather robust. Here, only the amplitude of Langmuir wave potential is changed and other parameters keep the same as discussed above.

When the amplitude is getting larger, the number of trapped electrons also increases, leading to an increase of the frequency shift.\cite{Berger2} The amplitude dependence of positions of the brightest sideband signals in $k_{x}$-$k_{y}$ space are shown in Fig~\ref{gamma_s} (a). The $|k_{x}-k_{L}|$ for oblique sidebands increases with the electrostatic fields, which is similar to the longitudinal sidebands. And the maximum value of $k_{y}$ for oblique sidebands also increases with the electrostatic fields, which is similar to the transverse sidebands.

In Fig.~\ref{gamma_s} (b), we extract the maximum growth rates of longitudinal sidebands (red circles) and the growth rate of the oblique sideband  with $k_{y} = 0.03125$ (blue squares). Especially, the dependence on the  amplitude for longitudinal sideband is obtained by power law fitting $\gamma = g_{1}\phi^{g_{2}}$ with $g_{2} = 0.75$, and that for oblique sideband is $g_{2} = 0.85$. Our scaling law for longitudinal sideband agree with early work $g_{2} \sim 0.6-0.9$.\cite{Friou} $g_{2}$ for sidebands are close to $0.5$, which reminds us the sideband instability has a strong relationship with the trapped-particle induced nonlinear frequency shift, which is also the starting point of our theoretical model of multidimensional sideband instability.

\section{Theoretical Model: sidebands of Langmuir waves  }\label{theory model}

The Dewar-Rose-Yin (DRY) model was first used to study the 1D modulation instability of Langmuir wave and then extended to study 2D modulation and filamentation instabilities.\cite{Dewar,rose1,rose2,Berger2} Here we reuse this model to predict the sidebands of Langmuir wave. In 2D situation, the modulation instability can be described by nonlinear Schr\"{o}dinger equation,
 \begin{equation}\label{dry1}
 \begin{split}
  i&\left(\frac{\partial}{\partial t}+\frac{\partial \omega_{k}}{\partial k_{x}} \frac{\partial}{\partial x}+\upsilon\right)\phi_{k}\\
   &+ \left(\frac{1}{2}\frac{\partial^{2} \omega_{k}}{\partial k_{y}^{2}} \frac{\partial^{2}}{\partial y^{2}}+\frac{1}{2}\frac{\partial^{2} \omega_{k}}{\partial k_{x}^{2}}\frac{\partial^{2}}{\partial x^{2}} + \Delta\omega\right)\phi_{k} = 0,
 \end{split}
 \end{equation}
  where $\phi_{k}(x,y,t)$ is the envelope of potential, $\nu$ is the Landau damping, and $\omega_{k}$ is the root of the dispersion relation $\epsilon(\vec{k},\omega) = 0$. The fluid type of dispersion relation is used,\begin{equation}\label{dry2}
  \epsilon(\vec{k},\omega) = \epsilon(k,\omega) = 1-\frac{\omega_{pe}^{2}}{\omega^{2}-3k^{2}v_{te}^{2}},
 \end{equation} where $k = |\vec{k}|=\sqrt{k_{x}^{2}+k_{y}^{2}}$ is the wavenumber of sidebands. The nonlinear term causing instability is the trapped-particle induced nonlinear frequency shift, $\Delta\omega$. To express it analytically, we use the adiabatic driven formula, \cite{Berger1}
 \begin{equation}\label{dry3}
  \frac{\Delta\omega}{\omega_{pe}}=-\frac{\alpha}{\sqrt{2\pi}(k_{L}\lambda_{De})^{2}}\sqrt{\frac{eE_{max}}{T_{e}k_L}}(v^{2}-1)exp(-\frac{v^{2}}{2})|_{v = v_{\phi}/v_{te}}.
\end{equation}
where $\alpha=0.544$, $v_{\phi}$ is the phase velocity of Langmuir wave and $E_{max} = 0.2 m_{e}\omega_{pe}v_{te}/e$ is the electron field amplitude of the Langmuir wave, which corresponds to a potential of $|e\phi/T_{e}| = 0.6$.

In order to find instability growth rate of Eq.~\ref{dry1},  we assume two small sidebands $\omega_{k} = \omega_{L}\pm\Omega$, $k_{x} = k_{L}\pm \Delta k_{x}$ and $k_{y} = k_{Ly}\pm \Delta k_{y}$ where $\omega_{L}$, $k_{L}$ are linear frequency and wavenumber of Langmuir wave, and in our consideration, $k_{Ly} = 0$. Let the Landau damping $\nu$ be absorbed by $\Omega$. Assuming a trial solution $\phi(x,y,t)=1/2(\phi_{k}(x,y,t)exp(i\vec{k}\cdot \vec{x} - i\omega_{k}t )+ c.c.)$ for  Eq.~\ref{dry1}, we can obtain the dispersion relation,
\begin{equation}\label{dry4}
\begin{split}
  &\left(\Omega - \Delta k_{x}\frac{\partial\omega_{k}}{\partial k_{x}}\right)^{2}=\left(\frac{\Delta k_{x}^{2}}{2}\frac{\partial^{2} \omega_{k}}{\partial k_{x}^{2}}+\frac{\Delta k_{y}^{2}}{2}\frac{\partial^{2} \omega_{k}}{\partial k_{y}^{2}}\right)\\
  &\left(\frac{\Delta k_{x}^{2}}{2}\frac{\partial^{2} \omega_{k}}{\partial k_{x}^{2}}+\frac{\Delta k_{y}^{2}}{2}\frac{\partial^{2} \omega_{k}}{\partial k_{y}^{2}}+\frac{\Delta\omega}{2}\right),
\end{split}
\end{equation}
where $\Delta k_{x} = k_{x} - k_{L}$, $\Delta k_{y} = k_{y}$. The growth rate of sideband is
\begin{equation}\label{dry5}
 \gamma(k_{x},k_{y}) = Im(\Omega)+\upsilon.
\end{equation} \begin{figure}[tbp]
    \begin{center}
      \includegraphics[width=0.49\textwidth,clip,angle=0]{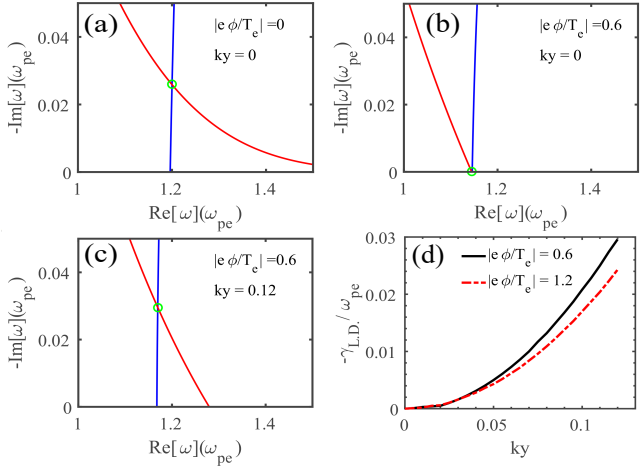}
      \caption{\label{damp}(a) Contours of solutions to the Langmuir wave when $|e\phi/T_{e}| = 0$ and $k_{y} = 0$, blue line stands for the real part of dielectric function, red line is the imaginary part of dielectric function. (b) Contours of solutions to the Langmuir wave when $|e\phi/T_{e}| = 0.6$ and $k_{y} = 0$. (c) Contours of solutions to the Langmuir wave when $|e\phi/T_{e}| = 0.6$ and $k_{y} = 0.12$. (d) The Landau damping as a function of $k_{y}$ with different amplitudes.}
    \end{center}
  \end{figure}The imaginary part of $\Omega$ exists only when $(\Delta k_{x}^{2}\frac{\partial^{2} \omega_{k}}{\partial k_{x}^{2}}+\Delta k_{y}^{2}\frac{\partial^{2} \omega_{k}}{\partial k_{y}^{2}})\Delta\omega<0$. Since $\frac{\partial^{2} \omega_{k}}{\partial k_{x}^{2}} =\frac{\partial^{2} \omega_{k}}{\partial k_{y}^{2}} \approx 3v_{te}^{2}/\omega_{L}$, thus modulation instability exists only when $\Delta\omega<0$. The maximum value of $Im(\Omega)$ and its location are easily solved from Eq. (\ref{dry4}),
\begin{eqnarray}
&(Im(\Omega))_m=\frac{\Delta\omega}{4}, \\
&(k_{x} - k_{L})^{2}+k_{y}^{2}=\frac{\Delta\omega\omega_L}{6v_{te}^2}.
\end{eqnarray}
Since $\Delta\omega$ is independent of $k$, if no damping is considered $\nu=0$, a constant maximum growth rate is located along a circle of the initial Langmuir wavenumber, $(k_{L}, 0)$. This is, of course, not consistent with our simulation results, so we must consider the distribution of Landau damping, $\upsilon=\upsilon(k_x,k_y)$.

\begin{figure}[tbp]
    \begin{center}
      \includegraphics[width=0.5\textwidth,clip,angle=0]{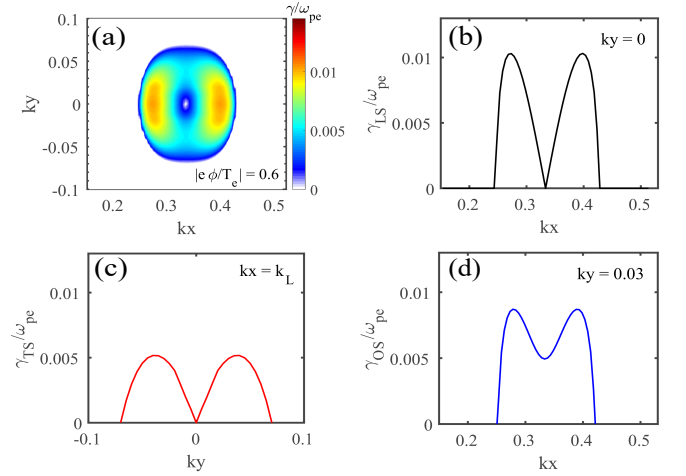}
      \caption{\label{theory_side}(a) The growth rates of sidebands obtained by DRY model, $|e\phi/T_{e}| = 0.6$. (b) The growth rates of longitudinal sidebands. (c) The growth rates of transverse sidebands. (d) The growth rates of oblique sidebands with $k_{y} = 0.03$.}
    \end{center}
  \end{figure}

 To do so, we must calculate the Landau damping in a trapping distribution function, such as the one presented at $t=200\omega_{pe}^{-1}$ in our simulation. Following the treatment of early works,\cite{Berger2,Divol} a two-dimensional trapping distribution function is artificially constructed,
 \begin{equation}\label{disfun}
\begin{split}
 &f_{0}(v_{x},v_{y}) = N_{e}f_{0,x}(v_{x})f_{0,y}(v_{y}),\\
 &f_{0,y}(v_{y}) = f_{M}(v_{y}),\\
 &f_{0,x}(v_{x}) = f_{M}(v_{x})+\delta f_{x}^{flat}(u),\\
 &\delta f_{x}^{flat}(u) = P(u)  \frac{1}{\sqrt{2}v_{te}} exp(-u^{2}/2),\\
 & P(u) = \beta u+\gamma(u^{2}-1),\\
 & u = (v_{x} - v_{\phi})/\delta v,
\end{split}
\end{equation}
where $f_{M}(v_{x})$ and $f_{M}(v_{y})$ are one-dimensional Maxwellian distributions. $\delta v = 2\sqrt{e\phi/T_{e}}$ is the width of the plateau around the phase velocity. $\beta$ and $\gamma$ are the first and second derivatives of $f_{0,x}(v_{x})$ at phase velocity, \begin{equation}\label{disfun1}
\begin{split}
 &\beta = (\delta \tilde{v})\tilde{v}exp(-\tilde{v}^{2}/2)|_{\tilde{v}=v_{\phi}/v_{te}},\\
 &\gamma = \frac{\delta \tilde{v}}{3}(1-\tilde{v}^{2})exp(-\tilde{v}^{2}/2)|_{\tilde{v}=v_{\phi}/v_{te}},\\
\end{split}
\end{equation}
where $\delta \tilde{v} = \delta v / v_{te}$. The constructed distribution function resembles Fig.~\ref{vx_vy}(a), which enables the zero Landau damping near the phase velocity $v_\phi=\omega_L/k_L$. The kinetic dispersion function under the non-Maxwellian distribution function is
\begin{equation}\label{KineticDisp}
\epsilon(\omega,k)=1-\frac{\omega_{pe}^{2}}{k^{2}}P\int^{\infty}_{-\infty}\frac{\nabla_v f_0\cdot d\textbf{v}}{v-\omega/k}-i\pi\frac{\omega_{pe}^{2}}{k^{2}}|\nabla_v f_0|_{v=\frac{\omega}{k}},
\end{equation}
where $P$ means the Cauchy principal value, $\textbf{v}=(v_x,v_y)$, and $k=\sqrt{k_x^2+k_y^2}$.

  \begin{figure}[tbp]
    \begin{center}
      \includegraphics[width=0.5\textwidth,clip,angle=0]{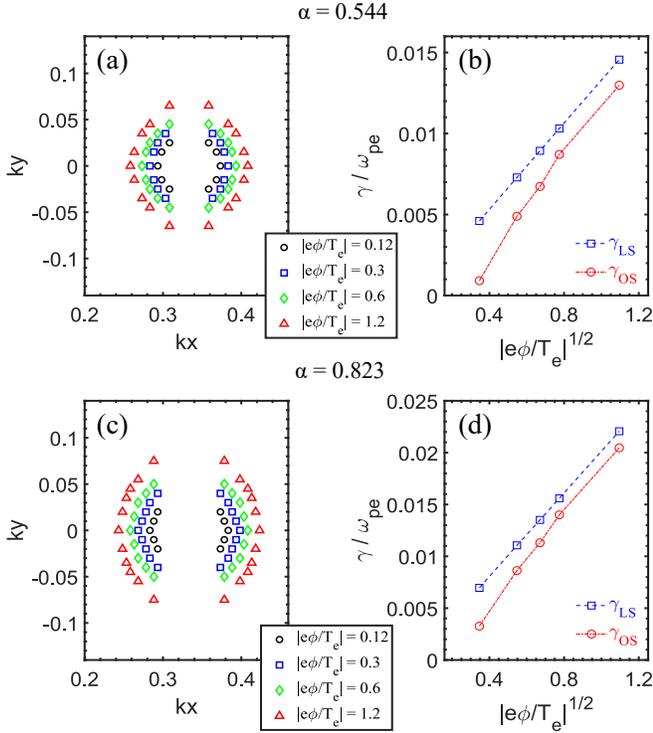}
      \caption{\label{theory_all}Theoretical results obtained by DRY  model. (a) and (b) use $\alpha = 0.544$, which corresponds to adiabatic approximation.\cite{Berger1} (a) The positions of oblique sidebands for different amplitude of field at $k_{x}$-$k_{y}$ space. (b) The amplitude of field dependence of growth rate of longitudinal sideband with maximum growth rate(blue squares) and the growth rates of oblique sideband with $k_{y} = 0.03$. (c) and (d) are the same results by using $\alpha = 0.823$, which corresponds to sudden approximation.\cite{Berger1}}
    \end{center}
  \end{figure}

Since the distribution function is almost flattened near the Langmuir wave phase velocity but decreases dramatically when $k_y$ increases, we assume the Landau damping varies with $k_{y}$ but is constant along $\Delta k_{x}$. Then we can numerically solve the dispersion function Eq. (\ref{KineticDisp}) along $k_x=k_L$ and obtain the Landau damping as a function of $k_y$ in Fig.~\ref{damp}. When $|e\phi/T_{e}| = 0$, it backs to the Maxwellian distribution function, so the Landau damping is $-0.026 \omega_{pe}$ as shown in Fig.~\ref{damp}(a). Fig.~\ref{damp}(b) shows the Landau damping at $k_{y} = 0$ when $e\phi/T_{e} = 0.6$, and the evaluated Landau damping being $0$ satisfies our designed distribution function. As $k_{y}$ increases to $0.12$ in Fig.~\ref{damp}(c), the phase velocity of sidebands becomes away from the trapped plateau, so the Landau damping increases to $-0.0296 \omega_{pe}$. Fig.~\ref{damp}(d) shows the overall Landau damping increases with $k_y$. And the Landau damping will decrease with the potential, because when $|e\phi/T_{e}|$ increases from $0.6$ to $1.2$, the width of trapped region becomes larger in accordance with our expectation. We should note that the distribution in Eqs.~\ref{disfun} is  artifical, thus, $v_{\phi}$ in Eqs.~\ref{disfun} and Eqs.~\ref{disfun1} is adjustable to make sure the Landau damping equals to $0$, when$ |e\phi/T_{e}| \neq 0$ and $k_{y} = 0$.

\begin{figure}[tbp]
    \begin{center}
      \includegraphics[width=0.48\textwidth,clip,angle=0]{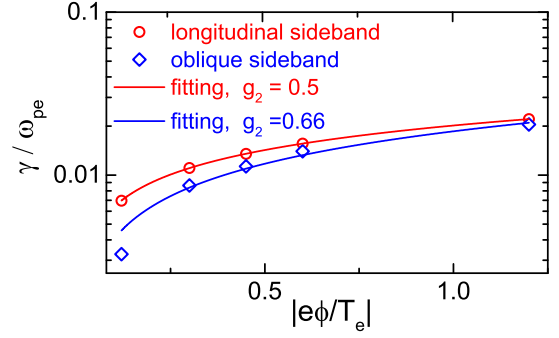}
      \caption{\label{fit_t}Theoretical results by DRY model for $k\lambda_{De} = 1/3$ with sudden approximation. The scaling law of growth rates of sidebands are obtained by using power law fitting $\gamma = g_{1}\phi^{g_{2}}$. We find that $g_{2} = 0.5$ for longitudinal sideband and $g_{2} = 0.66$ for oblique sideband with $k_{y} = 0.03$.}
    \end{center}
  \end{figure}
Fig.~\ref{theory_side} shows the final theoretical growth rates of sidebands obtained when $|e\phi/T_{e}|  = 0.6$. The shape of sidebands in Fig.~\ref{theory_side}(a) resembles the spectrum in Fig.~\ref{sideband}(c). We pick out three kinds of sidebands in Fig.~\ref{theory_side}(b), (c) and (d). Fig.~\ref{theory_side}(b) represents the longitudinal sidebands, where its maximum growth rate locates at $k_{x} = 0.2733, k_{y} = 0$ and the maximum growth rate is $\gamma_{LS} = 0.0103 \omega_{pe}$. In our simulations, the corresponding quantities are $k_{x} = 0.25, k_{y} = 0$, and $\gamma_{LS} = 0.02199 \omega_{pe}$. The DRY model underestimates $\delta_{k}$ and the growth rate of sideband perhaps due to the underestimation of the nonlinear frequency shift $\Delta \omega$ in Eq.~\ref{dry3}. Fig.~\ref{theory_side}(c) is the transverse sidebands with $k_{x} = k_{L}$. The position of transverse sideband with the maximum growth rate locates at $k_{x} = k_{L}, k_{y} = 0.04$, and the maximum growth rate of transverse sideband is $\gamma_{TS} = 0.0052 \omega_{pe}$. While in simulations they are $k_{x} = k_{L}, k_{y} = 0.04167$ and $\gamma_{TS} = 0.00588 \omega_{pe}$, respectively, which show good consistencies for transverse sidebands. This is because the transverse sidebands are mostly affected by Landau damping of sidebands,\cite{Berger2} showing the Landau damping in our calculation is accurate. Fig.~\ref{theory_side}(d) stands for the growth rates of oblique sidebands with $k_{y} = 0.03$, which is between the longitudinal and transverse sidebands as indicated by our Vlasov simulations.

\begin{figure}
    \begin{minipage}[t]{0.5\linewidth}
    \centering
    \includegraphics[width=1.8in]{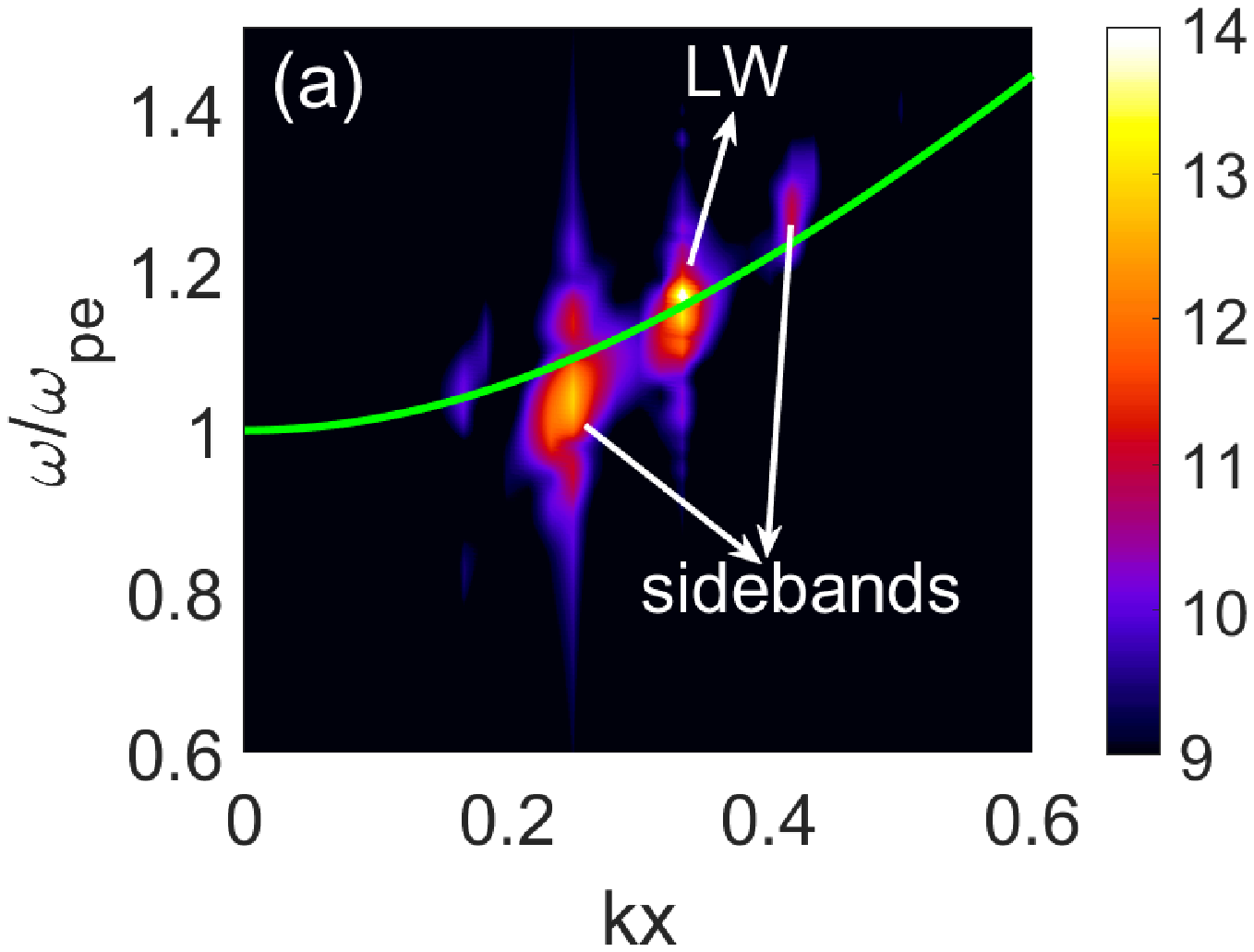}
    \label{fig:side:a}
    \end{minipage}%
    \begin{minipage}[t]{0.5\linewidth}
    \centering
    \includegraphics[width=1.8in]{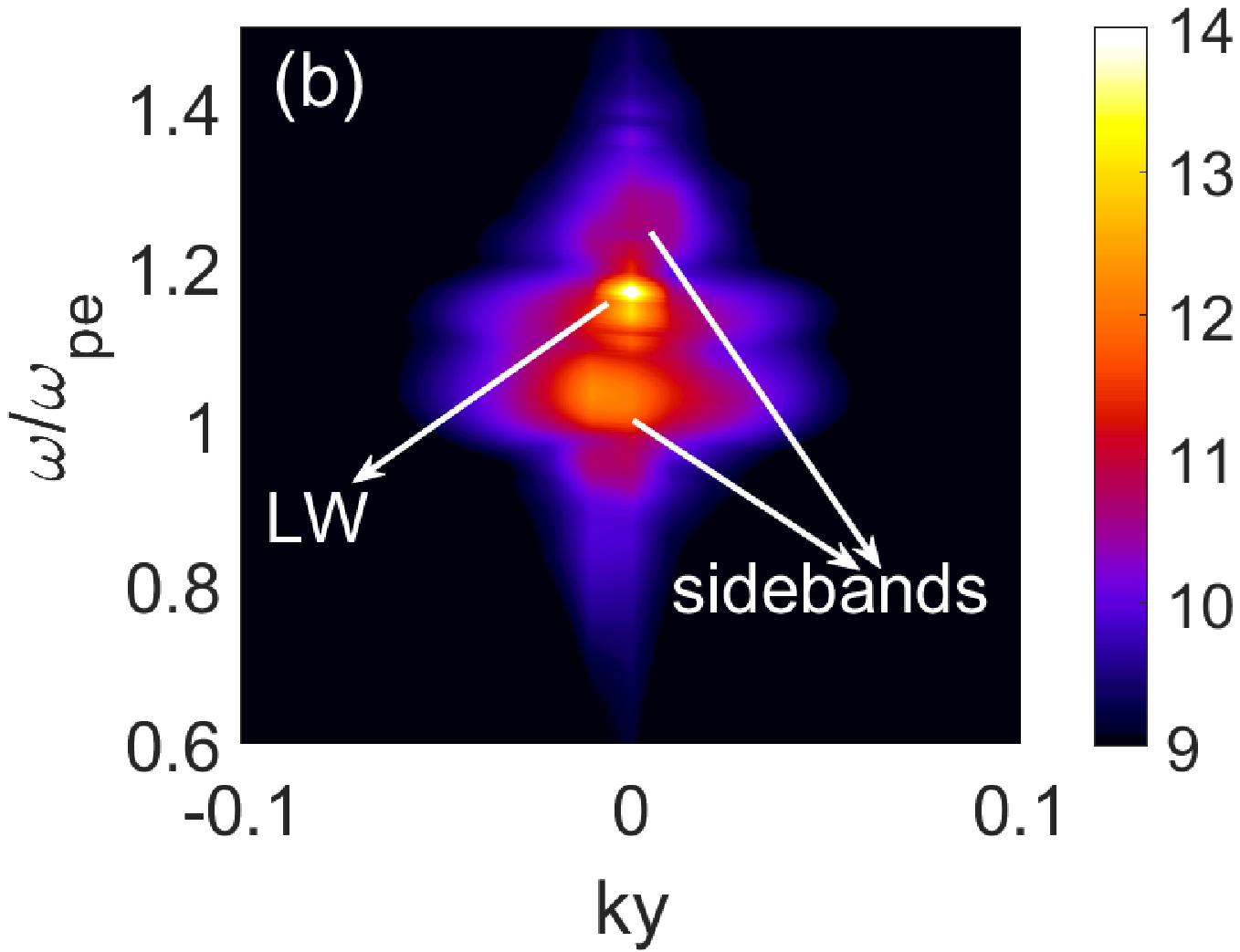}
    \label{fig:side:b}
    \end{minipage}
    \caption{\label{theory_3}Vlasov simulation results with $|e\phi/T_{e}| = 0.6$. (a) The spectrum of $E_{x}$ in $k_{x}$-$\omega$ space with a time window  $t\omega_{pe} = 0\sim1000$. The green line represents the dispersion relation of Langmuir wave.  (b) The spectrum of $E_{x}$ in $k_{y}$-$\omega$ space with a time window $t\omega_{pe} = 0\sim1000$.}
\end{figure}

The theoretical shapes of sidebands with the maximum growth rates and the corresponding growth rates as a function of potential amplitude are shown in Fig.~\ref{theory_all}. To compare with the Vlasov results, a sudden driven model of nonlinear frequency shift is plotted in Fig.~\ref{theory_all}(c) and (d), in addition to the adiabatic model shown in Fig.~\ref{theory_all}(a) and (b). The difference is that in a sudden driven model, $\alpha = 0.823$ in Eq.~\ref{dry3},\cite{Berger1} which permits a higher Langmuir wave frequency shift. Fig.~\ref{theory_all}(a) shows the amplitude dependence of the positions of sidebands. As compared with Fig.~\ref{gamma_s}(a), the shape and dependence on amplitude are similar, but it shrinks inward. When turning to the sudden driven model with $\Delta\omega_{sudden}\approx1.5\Delta\omega_{adiabatic}$, the shapes get large. In Fig.~\ref{theory_all}(b) and (d), the growth rates of longitudinal sidebands scale linearly with $|e\phi/T_{e}|^{1/2}$, which agrees with early theoretical results.\cite{Dewar,Berger2} As the nonlinear frequency shift gets larger, Fig.~\ref{theory_all}(d) shows a better agreement with the simulation results in Fig.~\ref{gamma_s}(b). In Fig.~\ref{fit_t}, we obtain the scaling law for sidebands by power law fitting under sudden approximation. $g_{2} = 0.5$ for longitudinal sideband and $g_{2} = 0.66$ for oblique sideband with $k_{y} = 0.03$. Both in simulation and theoretical results, $g_{2}$ for oblique sideband is a little larger than that of longitudinal sidebands. This may be because the influence of Landau damping on oblique sidebands.

However, there still exist discrepancies between Vlasov simulations and DRY model, especially when $|e\phi/T_{e}|$ is large. We suspect that it is due to the underestimation of $\Delta \omega$ in DRY model. Therefore, in Fig~\ref{theory_3}(a) and (b), we plot  the spectra of $E_{x}$ in the $k_{x}$-$\omega$ space and the $k_{y}$-$\omega$ from Vlasov simulations with $|e\phi/T_{e}| = 0.6$, respectively. The observed nonlinear frequency shifts of Langmuir wave and sidebands is on the order of $\Delta \omega \approx 0.1 \omega_{pe}$, however, theory predicts that $\Delta \omega_{adiabatic} = 0.0413 \omega_{pe}$ for adiabatic approximation and $\Delta \omega_{sudden} = 0.0624 \omega_{pe}$ for sudden approximation. Though with some flaws, the DRY model is good enough to capture the main physics of sidebands of Langmuir waves, and the growth rates of sidebands are qualitatively agreed with our 2D electrostatic Vlasov simulations.

\section{Conclusion and discussion}\label{conclusion}

In this paper, first, we study the evolution of Langmuir waves in a muti-wavelength system through 2D Vlasov simulations. There are three stages in our simulations. In the drive phase, an external driver is used to excited a monochromatic Langmuir waves. In the phase of sideband growth, the sidebands of Langmuir waves are excited because of the trapped electrons. Three kinds of sidebands are observed, which are the longitudinal sidebands, oblique sidebands and transverse sidebands. The oblique sidebands are arc-shaped in the k-space. The growth rates of oblique sidebands are smaller than that of the longitudinal sideband with the maximum growth rate but higher than the transverse sideband. In the turbulent state, the vortex merging happens when the amplitudes of sidebands are comparable with that of Langmuir waves. After vortex-merging, $20$ vortices merges to $15$ vortices. And The spectra of $k_{x}$ and $k_{y}$ become broader, inducing the filamentation of Langmuir waves. The amplitude dependence of sidebands are also studied by Vlasov simulations. We find that $\Delta k_{x}$ of longitudinal sidebands and $\Delta k_{x}$ and $\Delta k_{y}$ of oblique sidebands increase with the amplitude. Next, we reuse the DRY model to study the sidebands of Langmuir waves. Based on the early work,\cite{Berger2} the Landau damping with $k_{y}$ should be considered. The growth rates of three kinds of sidebands are finally obtained by DRY model, and they are qualitatively agreed with our Vlasov simulations.

The sideband instability is a saturation mechanism of SRS in ICF, and in earlier works, researchers mostly focused on the Longitudinal sidebands of Langmuir wave.\cite{Brunner1,Friou} The Vlasov simulations and theoretical results show clearly that sideband of Langmuir wave is a multidimensional instability. So, the sidebands of Langmuir wave in oblique and tranverse direction should also be considered. In this paper, we only consider the situation in two dimensions. The sideband with $k_{z}$ component might also exist. Besides, in early works, some kinetic effects by particle trapping have been studied by using particle in cell (PIC) simulations, such as Langmuir wave bowing\cite{yin,Lin} the trapped particle side loss \cite{hxvu3} in multidimensions. These works indicate that Multidimensional effects are important for instabilities in ICF.  In the future works, PIC simulations will be used to study the sidebands of Langmuir wave in multidimensions.

\section*{Acknowledgements}
We are pleased to acknowledge useful discussions with T. Yang, Y. Z. Zhou and S. X. Xie.  This work was supported by the Strategic Priority Research Program of Chinese Academy of Sciences (Grant No. XDA25050700), National Natural Science Foundation of China (Grant Nos. 11805062, 11875091 and 11975059), Science Challenge Project, No. TZ2016005 and Natural Science Foundation of Hunan Province, China (Grant No. 2020JJ5029).

\section*{DATA AVAILABILITY}
The data that support the findings of this study are available from the corresponding author upon reasonable request.


\end{document}